\documentclass[lettersize,journal]{IEEEtran}
\usepackage{amsmath,amssymb,amsfonts}
\usepackage{algorithmic}
\usepackage{algorithm}
\usepackage{array}
\usepackage[caption=false,font=normalsize,labelfont=sf,textfont=sf]{subfig}
\usepackage{textcomp}
\usepackage{stfloats}
\usepackage{siunitx}
\usepackage{soul}
\usepackage{url}
\usepackage{hyperref}
\hypersetup{
	colorlinks=true,
	linkcolor=blue,  
	urlcolor=blue,
	citecolor=blue
}
\usepackage{verbatim}
\usepackage{graphicx}
\usepackage{cite}
\usepackage{orcidlink}

\begin{document}

\title{Physics-Based and Closed-Form Model for \\ Cryo-CMOS Subthreshold Swing}

\author{Arnout Beckers\,\orcidlink{0000-0003-3663-0824}, Jakob Michl\,\orcidlink{0000-0003-2539-3245}, Alexander Grill\,\orcidlink{0000-0003-1615-1033}, Ben Kaczer\,\orcidlink{0000-0002-1484-4007}, Marie Garcia Bardon\,\orcidlink{0000-0001-5772-5406}, Bertrand Parvais\,\orcidlink{0000-0003-0769-7069}, Bogdan Govoreanu\,\orcidlink{0000-0001-7210-2979}, Kristiaan De Greve\,\orcidlink{0000-0002-1314-9715}, Gaspard Hiblot\,\orcidlink{0000-0002-3869-965X} and Geert Hellings\,\orcidlink{0000-0002-5376-2119}

\thanks{The authors are with imec, Leuven, Belgium (arnout.beckers@imec.be). B. Parvais is also with the Vrije Universiteit Brussel, Belgium. K. De Greve is also with KU Leuven. J. Michl was with the Institute for Microelectronics, TU Vienna, Austria, at the time of initial submission.}
}
% The paper headers
\markboth{Journal of \LaTeX\ Class Files,~Vol.~14, No.~8, February~2023}%
{Beckers \MakeLowercase{\textit{et al.}}: Physics-Based and Closed-Form Model for Cryo-CMOS Subthreshold Swing}

\maketitle

\begin{abstract}
Cryogenic semiconductor device models are essential in designing control systems for quantum devices and in benchmarking the benefits of cryogenic cooling for high-performance computing. In particular, the saturation of subthreshold swing due to band tails is an important phenomenon to include in low-temperature analytical MOSFET models as it predicts theoretical lower bounds on the leakage power and supply voltage in tailored cryogenic CMOS technologies with tuned threshold voltages. Previous physics-based modeling required to evaluate functions with no closed-form solutions, defeating the purpose of fast and efficient model evaluation. Thus far, only the empirically proposed expressions are in closed form. This article bridges this gap by deriving a physics-based and closed-form model for the full saturating trend of the subthreshold swing from room down to low temperature. The proposed model is compared against experimental data taken on some long and short devices from a commercial 28-nm bulk CMOS technology down to 4.2\,K.
\end{abstract}


\section{Introduction}
\IEEEPARstart{I}{n} recent years, cryo-CMOS has emerged as the leading candidate for controlling quantum devices due to its proven performance down to deep-cryogenic temperatures and its potential for monolithic integration with quantum bits \cite{zalba,jazaeri_review_2019,sebastianocryocmos}. Other cutting-edge physics experiments, like gravitational wave or dark matter detection, can also benefit from the increased attention for cryo-CMOS devices and circuits \cite{tavernier_chip_2020,fahim}. Furthermore, datacenters could benefit from advanced CMOS technologies tailored to cryogenic operation \cite{chiang,arm2}. 

Steepness of the transistor switching between ON- and OFF-states at cryogenic temperatures is an essential performance metric for the aforementioned applications. In a MOSFET, the inverse slope of the subthreshold current in $\log_{10}$-scale, a.k.a. subthreshold swing, $SS$, is expected to scale linearly with temperature ($T$), following Boltzmann's thermal limit, 
\begin{equation}
	SS_{\,(T\gg T_c)}=m\cdot \frac{k_BT}{q}\cdot \ln(10),
	\label{eq:ht}
\end{equation}
where $m$ is the subthreshold-slope factor, $k_B$ is Boltzmann's constant, and $q$ is the electron charge. 

However, $SS$ deviates from this linear scaling around a certain critical temperature, $T_c$, typically around \SI{50}{\kelvin}. In previous literature \cite{bohus,edl,ghib}, $SS$ has been shown to saturate to a temperature-independent plateau below $T_c$, 
\begin{equation}
	SS_{\,(T\ll T_c)}=m\cdot \frac{k_BT_c}{q}\cdot \ln(10),
	\label{eq:lt}
\end{equation}
where $T_c$ links to the characteristic exponential decay of a band tail, i.e., $W_{t,c}=k_BT_c$. Such band tails have already been demonstrated experimentally using electron-spin resonance in silicon MOS devices at \SI{370}{\milli\kelvin} \cite{jock}. Interface and source/drain engineering give $SS$ closer to the thermal limit \cite{richstein_interface_2022,yihan}. Besides band tails, direct source-to-drain tunneling is an additional explanation for saturation of $SS$ at cryogenic temperatures in short devices \cite{frank,kaosourcedrain}.

Limits (\ref{eq:ht}) and (\ref{eq:lt}) are physics-based models, but they cover only part of the $T$-domain. To capture the full saturating trend across $T_c$, different models are being used.
The physics-based models numerically integrate the Fermi-Dirac integrals with exponential density-of-states \cite{ghib,bohus}, or use Gauss hypergeometric functions \cite{edl}, which are not of closed form and thus difficult to implement in compact models. On the other hand, suitable curve fits are available connecting the two limits, e.g., $\propto \ln(1+e^x)$ [(7), \cite{ghib}] or $\propto \sqrt{1+x^2}$ [(9), \cite{pahwa}]. 

From (\ref{eq:ht}) and (\ref{eq:lt}) it is clear that a $T_c/T$-dependence must appear out of the physics derivation to complete the transition below $T_c$ in a full $T$-range model. Such a $\propto 1/T$-dependence has been postulated since 1985 \cite{tewksbury} but is nowadays often inserted empirically in the Boltzmann limit with the help of an effective slope factor  \cite{akturk,enz_cryo-cmos_2020} or an effective temperature \cite{pahwa}. It has not been derived in a physical model thus far. The closest attempt were the discrete trap states close to the band edge in \cite{jeds} which gave a $\propto 1/T$-dependence at the state energy. This $\propto 1/T$ must also follow from a physics-based derivation for a continuous tail of states close to the band edge. In this article, such a $T_c/T$-dependence is derived inside a physics-based and closed-form model for $SS$ covering the whole temperature domain from $\approx$\,\SI{0}{\kelvin} to \SI{300}{\kelvin} (Section \ref{sec:derivation}). We explore the sensitivity of the physical model parameters and compare the model with experimental data (Section \ref{sec:exploration}).

\begin{figure}[t]
	\centering
	\includegraphics[width=0.5\textwidth]{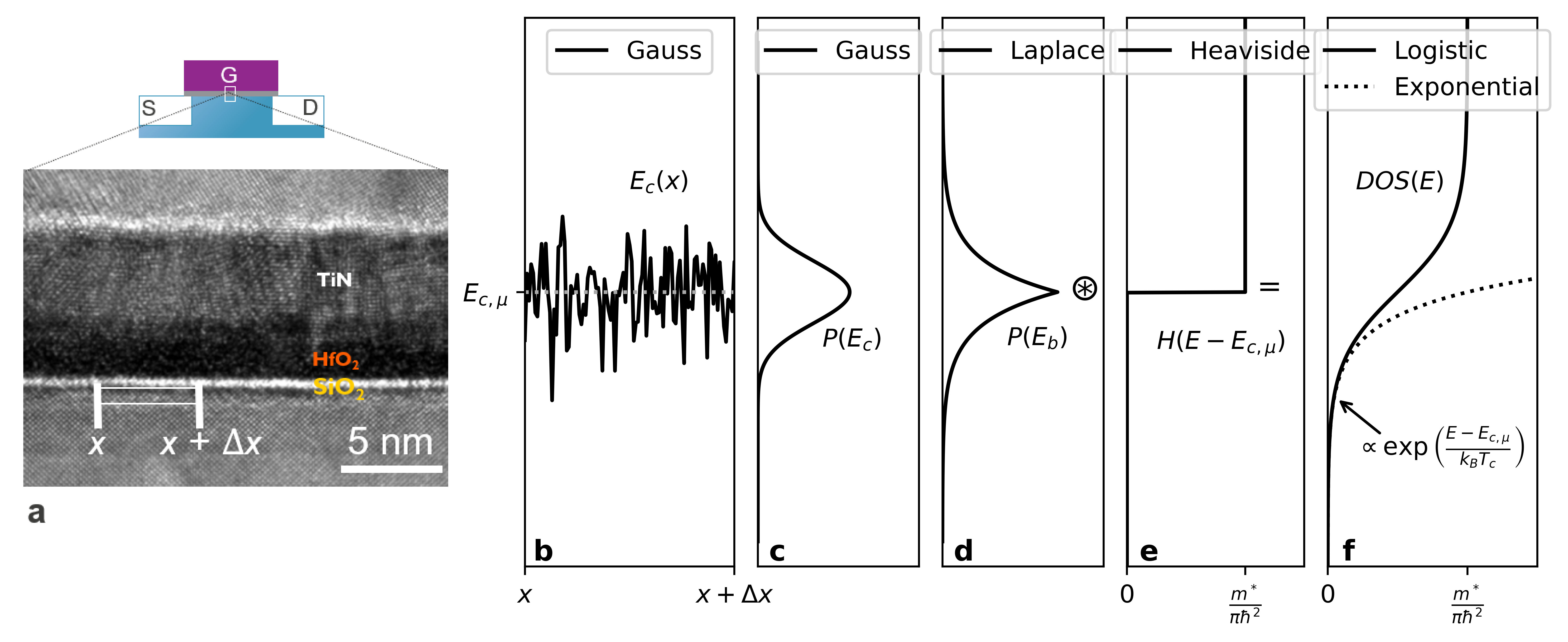}
	\vspace{-0.45cm}
	\caption{a) TEM picture of a mature imec technology node \cite{lars}, b) Electrostatic potential fluctuations near the channel/oxide interface, c) Gaussian distributed depths of the potential wells \cite{bacalis}. d) Including the binding energy in the wells in the quantum picture gives a Laplace distribution of $P(E_b)$ \cite{bacalis}. \textcolor{black}{e)-f) Convolution ($\circledast$) of $P(E_b)$ with the sharp-edged 2-D DOS leads to a logistic/Fermi-like DOS function with an exponential tail.}}
	\label{fig:randompotential}
\end{figure}

\section{\label{sec:derivation}Model Derivation \& Discussion}
\subsection{Exponential Band Tail}
Fig. \ref{fig:randompotential} clarifies, from left to right, the relation between the random potential in the channel of a MOSFET (assumed $n$-type) and the exponential tail in the density-of-states (DOS). Fig.\ref{fig:randompotential}(a) shows a TEM picture of an older imec technology node with $\mathrm{SiO}_2/\mathrm{HfO}_2/\mathrm{TiN}$ gate stack \cite{lars}. Fig.\ref{fig:randompotential}(b) shows the blurring of the conduction-band edge $E_c(x)$ along a segment of the channel between $x$ and $x + \Delta x$. 

Following \cite{bacalis}, the random potential in the channel is characterized by a Gaussian distribution. It is indeed reasonable to assume a Gaussian distribution since, according to the central limit theorem, the aggregate of a large number of independent random variables still has a normal distribution even if the variables are not normally distributed. The Gaussian distribution of the depths of the local potential wells in $E_c(x)$,
\begin{equation}
	P(E_c)\propto\exp\left(\frac{-(E_c-E_{c,\mu})^2}{2\sigma^2}\right)
	\label{eq:P}
\end{equation}
is centered around the average $E_{c,{\mu}}$ with a standard deviation of $\sigma$ [Fig. \ref{fig:randompotential}(c)]. To obtain the smeared-out DOS with band tailing, the probability distribution must be convolved with the sharp-edged DOS (see entry \textquotedblleft H\textquotedblright \, in Table \ref{table}), that is,
\begin{eqnarray}
	DOS(E)&=&  DOS_H \circledast P(E_c) \\
	&=&\int_{-\infty}^{E}DOS_H(E-E_{c})\cdot P(E_c)\cdot dE_c
	\label{eq:dosintegral}
\end{eqnarray}
where $H$ stands for the Heaviside step function and $\circledast$ stands for the convolution operation. 

\begin{figure}[t]
	\centering
	\includegraphics[width=0.49\textwidth]{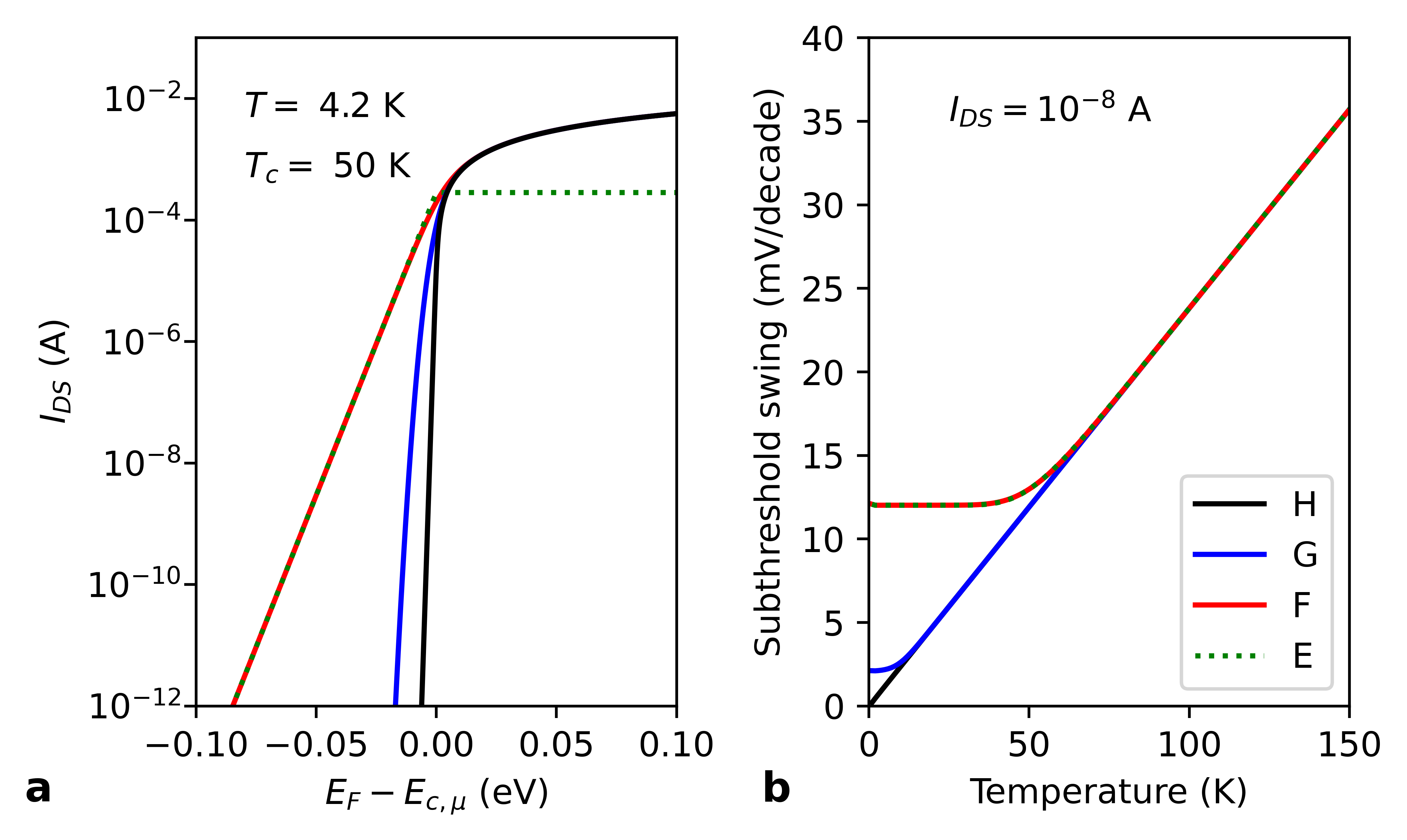}
	\caption{For $I_{DS}$ in subthreshold ($\approx \SI{e-8}{\ampere}$), the $SS$ obtained for a Fermi-like DOS (F : band tail + band) using numerical integration, falls on top of that for an exponential DOS (E, band tail only). Therefore, including only the exponential DOS, is a sufficiently accurate assumption for our present purposes of deriving a closed-form expression for $SS$. }
	\label{fig:ids}
\end{figure}

However, this convolution with a Gaussian distribution does not give rise to an exponential tail immediately. The resulting DOS would have an error function shape with tail $\propto \exp(-x^2)$ (see entry \textquotedblleft G\textquotedblright \, in Table \ref{table}). The exponential $DOS(E)$ arises out of the quantum picture, where the binding energies of the local potential wells in $E_c(x)$ are taken into account. Bacalis \emph{et al.} found an approximately linear relation between the binding energy ($E_b$) and the square of the depth of the potential well for different types of wells : $\vert E_b-E_{c,\mu}\vert \simeq A\cdot (E_c-E_{c,\mu})^2-B$ \cite{bacalis}. Inserting this expression in (\ref{eq:P}), we obtain a Laplace distribution for the binding energies 
\begin{equation}
	P(E_b)\propto\exp\left(\frac{-\vert E_b-E_{c,\mu}\vert}{k_BT_c}\right),
	\label{eq:PEb}
\end{equation}
shown in Fig.\ref{fig:randompotential}(d), where $k_BT_c=2A\sigma^2$. The convolution of the sharp DOS, shown in Fig.\ref{fig:randompotential}(e),  with (\ref{eq:PEb}) gives a logistic- or Fermi-like distribution plotted in Fig. \ref{fig:randompotential}(f). The dashed line shows the exponential behavior below $E_{c,\mu}$, which is often referred to as a \textquotedblleft band tail\textquotedblright. Note, in the semi-classical picture (i.e., no binding energies considered, hence no panel d in Fig.\ref{fig:randompotential}), \textquotedblleft G\textquotedblright \, would be obtained. The Gaussian tail gives a steeper subthreshold slope as shown in Fig. \ref{fig:ids}(a), due to a less fat tail since $\propto \exp(-x^2)$ goes faster to zero than $\exp(-x)$. Fig. \ref{fig:ids}(b) also checks that, for current values in subthreshold ($\approx 10^{-8}\si{\ampere}$), it is sufficient to take into account the exponential tail of the DOS only, without the band states. This allows to derive a closed-form model for $SS$ in the rest of the paper. 
\begin{table}[t]
	\centering
	\caption{Band-Tail Shapes}
	\label{table}
	\setlength{\tabcolsep}{3pt}
	\begin{tabular}{|p{25pt}|p{75pt}|p{115pt}|}
		\hline
		\textbf{Symbol}& \textbf{DOS shape}& \textbf{DOS expression} \par $x=\left(E-E_{c,\mu}\right)/k_BT_c$ \\
		\hline 
		& 
		\par & 
		\\
		$H$& 
		Standard \par (sharp, no band tail) \par & 
		$\frac{m^*}{\pi\hbar^2}\cdot H(x)$ \\
		$G$& 
		Normal (Gauss) \par & 
		$\frac{m^*}{\pi\hbar^2}\cdot 0.5 \cdot \left[\mathrm{erf}\left(x\right)+1\right] $\\
		$F$& 
		Logistic (Fermi) \par & 
		$\frac{m^*}{\pi\hbar^2} \cdot \left[1+\exp(-x) \right]^{-1}$ \\
		$E$& 
		Exponential (subthreshold approx.) \par & 
		$\frac{m^*}{\pi\hbar^2}\cdot \exp\left(x\right) \cdot H(-x)$\\
		\hline
	\end{tabular}
	\label{tab1}
\end{table}
\begin{figure*}[t]
	\centering
	\includegraphics[width=0.85\textwidth]{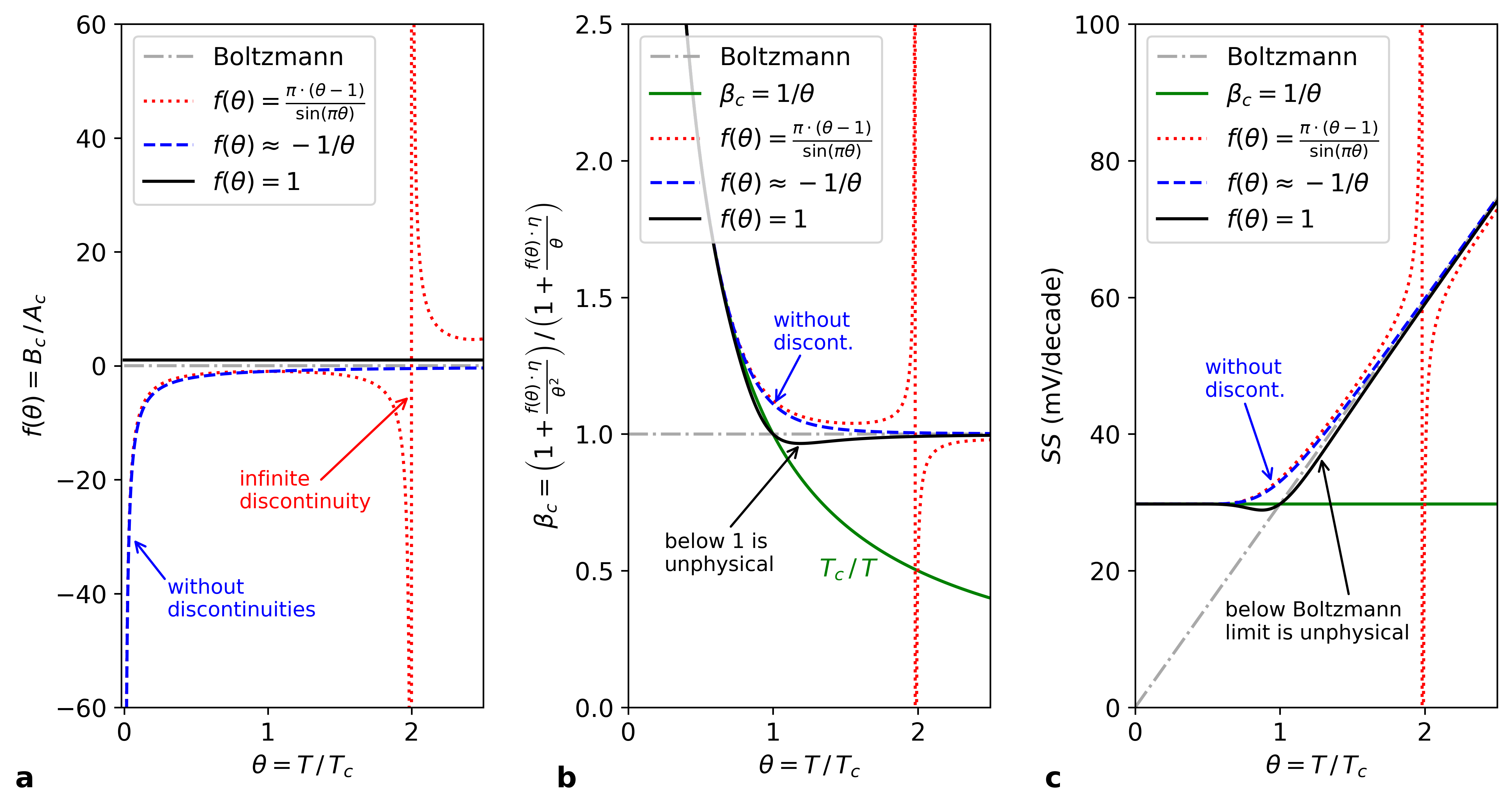}
	\vspace{-0.25cm}
	\caption{(a) Function $f(\theta)$ from (\ref{eq:f}) and approximations. The infinite discontinuities arise from the analytical solution of the Fermi-Dirac integral with exponential DOS. (b) The fraction $\beta_c$ from (\ref{eq:bc1}) for different approximations of $f(\theta)$. (c) $f(\theta)=1$ is a too simplistic approximation which leads to unphysical behavior in $SS$ around the transition temperature ($\theta=1$). The green lines in (b) and (c) indicate the low-temperature asymptote. The model shown with the blue dashed line will be used further in the rest of the paper.}
	\label{fig:bc2}
\end{figure*}
\subsection{Subthreshold Swing Derivation\label{sec:ss}}
The subthreshold swing is defined as 
\begin{equation}
	SS\triangleq\frac{\partial V_{GS}}{\partial \log_{10}(I_{DS})}=\frac{I_{DS}}{\frac{\partial I_{DS}}{\partial V_{GS}}}\cdot \ln(10),
	\label{eq:ssdef1}
\end{equation}
where $I_{DS}$ is the drain-source current in subthreshold and $V_{GS}$ is the gate-to-source voltage. The current flowing below $E_{c,\mu}$ would be accurately described by percolation theory \cite{cata}. The carriers percolate through the rugged potential landscape in Fig. \ref{fig:randompotential}(b).  \textcolor{black}{The band tail captures only the average effect of the potential fluctuations on the carriers traveling in a long channel. For short channels, tunneling and hopping currents need to be taken into account using the appropriate quantum formalism, see \cite{hungchi}. To obtain an expression for $I_{DS}$ in subthreshold, we approximate this carrier flow with drift-diffusion transport in first order.} We also adopt the \textcolor{black}{gradual-channel and} depletion approximations for obtaining the surface electric field from \textcolor{black}{the 1-D} MOS Poisson equation. 

The mobile sheet charge density is then given by 
\begin{equation}
	Q_i\approx -(m-1)\cdot C_{ox}\int_0^{\psi_s}\frac{n(\psi)}{N_A}\cdot d\psi
	\label{eq:Qi}
\end{equation}
where $C_{ox}$ is the gate-oxide capacitance, $\psi_s$ is the surface potential, and $N_A$ the doping concentration \cite{taur}. Recall that the Fermi-Dirac integral with exponential-tail DOS is 
\begin{equation}
	n=\int_{-\infty}^{E_c} DOS_E\left(\frac{E-E_c}{k_BT_c}\right)\cdot f_D(E)\cdot dE,
	\label{eq:int_n}
\end{equation}
where $f_D(E)$ is the Fermi-Dirac distribution function and $E_c$ is now the conduction-band edge at which the tail transitions into the band. This integral is exactly expressible as a Gauss hypergeometric (GH) function. Using mathematical formulas from Gradshteyn and Ryzhik for this special function \cite{gradshtein}, it is possible to derive a Boltzmann-like relation, which is valid in subthreshold. It features a sum of two exponentials: \cite{jap}
\begin{eqnarray}
	\label{eq:n}	n(\psi) &=& N_c \cdot A_c \cdot\exp\left(\frac{\psi-\Phi_{F,n}-0.5\cdot E_g}{U_T}\right)\\&+&N_c\cdot B_c\cdot\exp\left(\frac{\psi-\Phi_{F,n}-0.5\cdot E_g}{U_{T,c}}\right),\nonumber
\end{eqnarray}
where $N_c$ is an effective DOS, $U_T\triangleq k_BT/q$ is the thermal voltage, $U_{T,c}\triangleq k_BT_c/q$ is the \textquotedblleft band-tail thermal voltage\textquotedblright, $q\cdot\psi=E_c^o -E_c$ is the electrostatic potential, $q\cdot \Phi_{F,n}=E_c^o -E_{F,n}-0.5\cdot E_g$ is the quasi-Fermi potential of the electrons, $E_{F}$ is the Fermi level, and $E_c^{o}$ is the reference energy. The temperature-dependent coefficients in (\ref{eq:n}) are given by 
\begin{eqnarray}
	\begin{cases}
		\label{eq:A} \, A_c	=\theta/(\theta - 1),\\ 
		\, B_c=\pi\theta / \sin(\pi\theta),              
	\end{cases}
\end{eqnarray}
where $\theta\triangleq T/T_c$ is the operating temperature normalized to the critical temperature. The coefficients $A_c$ and $B_c$ will be important to obtain a well-behaved expression for $SS$ over $T$. The discontinuities of $A_c$ and $B_c$ at integer values of $\theta$ (multiples of $T_c$) do not necessarily point to a limitation in the derivation or in the physical assumptions. They can arise from the Gamma functions which are encountered in the mathematics of the Fermi-Dirac integrals with an exponential DOS \cite{jap}. \textcolor{black}{The expression for $I_{DS}$ is further derived in the Appendix, see (\ref{eq:p})-(\ref{eq:IDS}).} Differentiating (\ref{eq:IDS}) to $V_{GS}$ and plugging the result back into the definition of $SS$ (\ref{eq:ssdef1}), gives
\begin{align}
	SS=m\cdot \left(\frac{k_BT}{q}\right)\cdot \ln(10)\cdot \beta_c,
	\label{eq:ss}
\end{align}
where 
\begin{equation}
	\beta_c=\frac{1+f\cdot \eta/\theta^2}{1+ f\cdot \eta/\theta},  
	\label{eq:bc1}
\end{equation}
is a new multiplicative factor that contains the influence of the band tail. Recall that $\theta=T/T_c$ is the normalized temperature, and we have also defined the following functions
\begin{equation}
	f=\frac{B_c}{A_c}=\frac{\pi\cdot(\theta-1)}{\sin(\pi\theta)},
	\label{eq:f}
\end{equation}
\begin{equation}
	\eta=r\cdot e^{\frac{-V_{GT}}{m\cdot U_{T}}\cdot(1-\theta)},
	\label{eq:y}
\end{equation}
and 
\begin{equation}
	r=\left(1-e^{\frac{-V_{DS}}{U_{T,c}}}\right)/\left(1-e^{\frac{-V_{DS}}{U_{T}}}\right).
	\label{eq:r}
\end{equation}

Figs. \ref{fig:bc2}(a)-(c) plot $f(\theta)$, $\beta_c(\theta,\eta)$, and $SS$ versus $\theta$, respectively. At $\theta=2$, one of the infinite discontinuities appears, which arise at multiples of $\theta$ due to the $\sin$-function in the denominator of $f(\theta)$. We aim to overcome these discontinuities noting that $\eta$ goes fast to zero above $T_c$, therefore $f(\theta)$ is only important at low temperatures in the expression of $\beta_c$. The low-temperature asymptote $\left[\,\sin(\pi\theta)\approx \pi\theta\,\right]$ gives $f(\theta)\approx -1/\theta$, and therefore $\beta_c$ simplifies to
\begin{equation}
	\beta_c\approx\frac{1-\eta/\theta^3}{1-\eta/\theta^2},
	\label{eq:bc2}
\end{equation}
which does no longer give rise to infinite discontinuities at multiples of $\theta$ as shown with the dashed blue line in Fig.\ref{fig:bc2}(b). Fig.\ref{fig:bc2}(b) also shows that $\beta_c$ has the elusive $1/\theta=T_c/T$ asymptote at low temperatures that we have been searching for [green solid line]. This $1/T$ dependence has been postulated in early hypotheses but was not derived there \cite{tewksbury,akturk}. It emerges from the $I_{DS}$ factor instead of the slope factor. If the standard Boltzmann limit of $SS$, without $\beta_c$, is still used at cryogenic temperatures, an apparent $1/T$-dependence in $m$ will be extracted from the experimental $SS$ data \cite{jeds}. This can lead to unrealistic predictions of the interface-state density ($m\propto N_{it}$) \cite{galy}, and thus $\beta_c$ (or an equivalent expression) is best included. Parallels can be drawn between (\ref{eq:ss}) and the empirical models which insert an effective slope-factor\cite{tewksbury,akturk} or an effective temperature \cite{pahwa} inside the Boltzmann limit, by defining $m_{\mathrm{eff}}=m\cdot \beta_c$ or $T_{\mathrm{eff}}=T\cdot \beta_c$ in (\ref{eq:ss}). 

Note that $A_c$ and $B_c$ in (\ref{eq:n}) were important to obtain a well-behaved transition around $T_c$. Assuming [$A_c=B_c=1$ and thus $f(\theta)=1$], which corresponds to a standard Boltzmann relation for the band carriers, plus assuming a step-like Fermi function in the band tail, gives rise to an unphysical $SS(T)$ behavior dropping below Boltzmann's thermal limit around $\theta=1$, as shown with the black solid line in Fig. \ref{fig:bc2}(c).
\begin{figure}[t]
	\centering
	\includegraphics[width=0.49\textwidth]{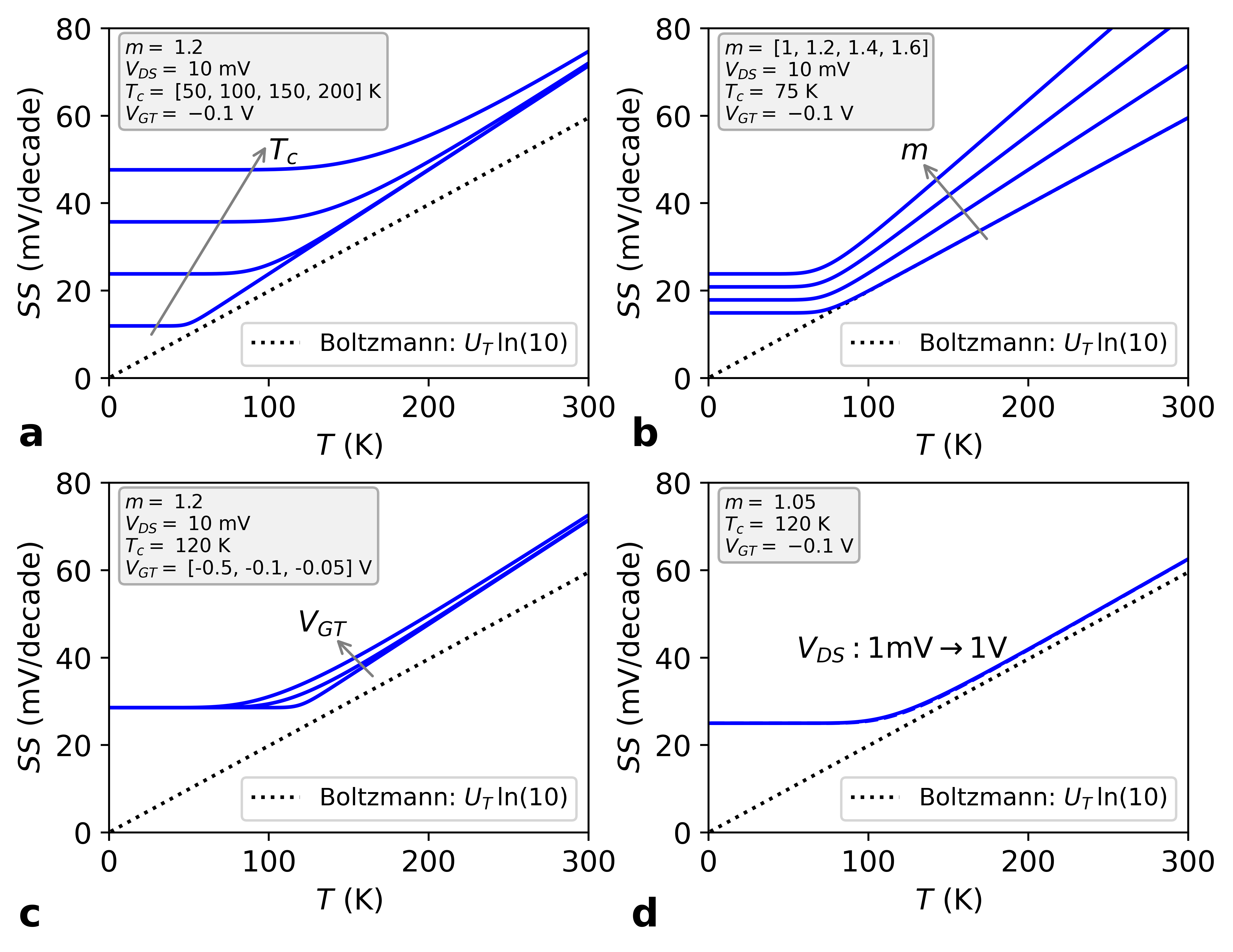}
	\vspace{-0.5cm}
	\caption{Sensitivity analysis using (\ref{eq:ss}) combined with (\ref{eq:y}), (\ref{eq:r}), and (\ref{eq:bc2}): \textcolor{black}{(a) Impact of increasing band-tail width $\propto T_c$, (b) Impact of increasing slope factor $m$, (c) The gate voltage has some impact on the transition between the two limits, and (d) No significant impact seen from the $V_{DS}$ dependence.}}
	\label{fig:sensitivity}
\end{figure}
\begin{figure}[t]
	\centering
	\includegraphics[width=0.5\textwidth]{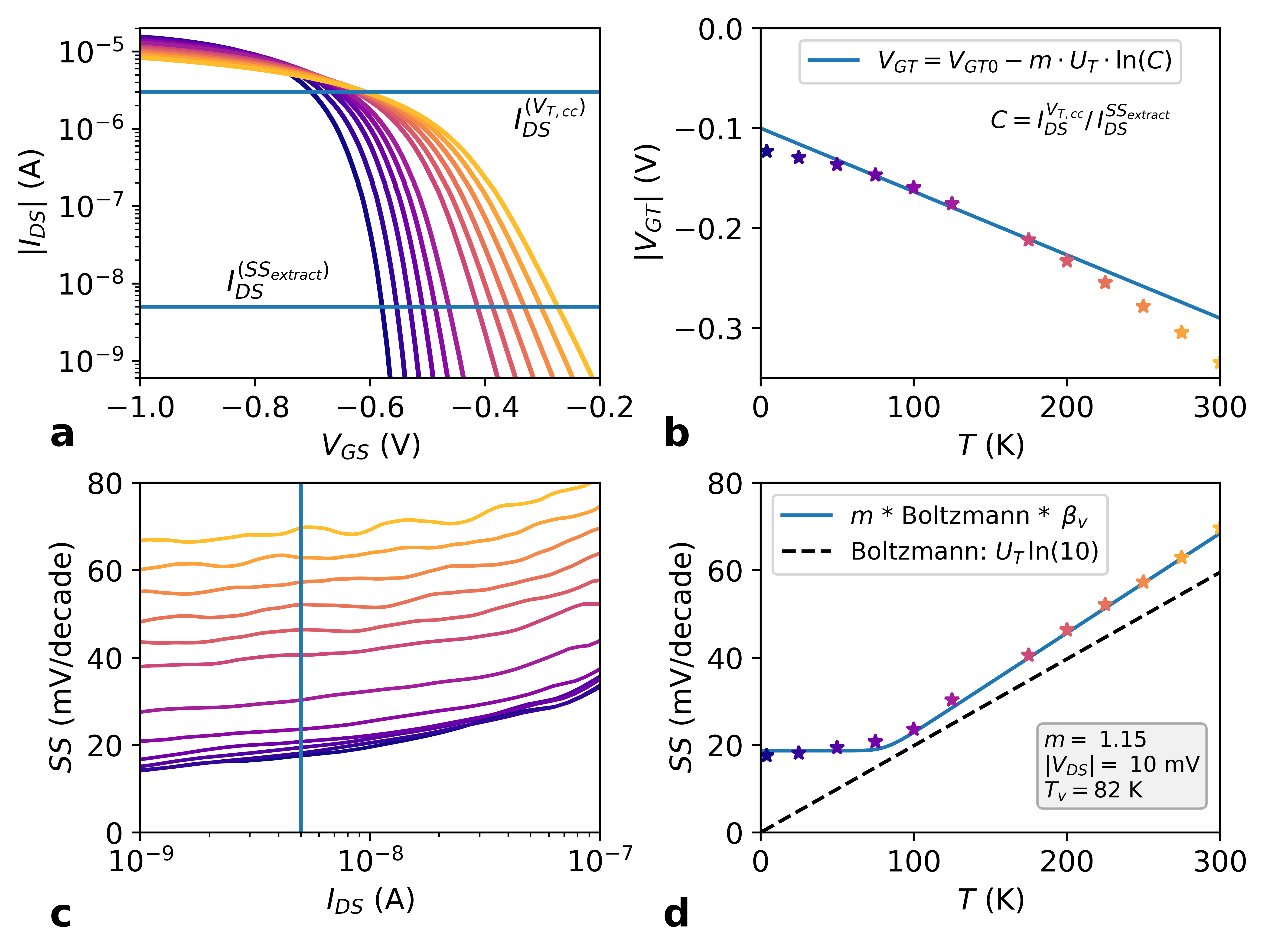}
	\vspace{-0.5cm}
	\caption{Subthreshold swing for PMOS $W/L=\SI{10}{\micro\meter}/\SI{1}{\micro\meter}$ from a 28-nm bulk CMOS process. (a) Transfer characteristics for $V_{DS}=\SI{10}{\milli\volt}$. (b) Difference between extracted gate voltage and threshold voltage at the two indicated current values. (c) Experimental $SS$ data. (d) $SS$ model fit.}
	\label{fig:ss_p}
\end{figure}

\section{\label{sec:exploration}Model Exploration, Verification, and Experimental Validation}
\subsection{Sensitivity of the Physical Model Parameters}
In this section, we use the full expression, i.e., (\ref{eq:ss}) combined with (\ref{eq:y}), (\ref{eq:r}), and (\ref{eq:bc2}), to explore the sensitivity to the physical model parameters $m$, $T_c$, $V_{GT}$, and $V_{DS}$. Fig. \ref{fig:sensitivity}(a) shows that increasing $T_c$ raises up the low-$T$ plateau, which is in line with the literature \cite{bohus,edl,ghib}. This allows to extract the band-tail decay ($W_{t,c}=k_BT_c$) from transport measurements at cryogenic temperatures. Fig. \ref{fig:sensitivity}(b) shows the impact of a larger slope factor due to bad gate control. The influence of $m$ on $SS$ is more pronounced in the high-$T$ limit.  

Fig. \ref{fig:sensitivity}(c) shows a smoother transition of $SS$ around $T_c$ for smaller $\vert V_{GT}\vert$ ($V_G$ closer to $V_T$ and $E_F$ closer to the band edge scanning a greater portion of the band tail). Thus $V_{GT}$ gives a (partial) physical explanation for the smoothing parameters $\alpha$ and $D_0$ introduced in the empirical closed-form expressions [(7), \cite{ghib}] and [(9), \cite{pahwa}], respectively. $V_{GT}$ is linked to the $I_{DS}$ value at which $SS$ is extracted from measurements ($I_{DS}^{SS, \mathrm{extract}}$). This will be used as part of the parameter extraction procedure for long devices in Section \ref{sec:comp}. Fig. \ref{fig:sensitivity}(d) shows the model for two different high and low $V_{DS}$ (\SI{1}{\milli\volt} up to \SI{1}{\volt}). No noticeable difference is obtained. %when varying $V_{DS}$.
\subsection{\label{sec:comp}Extraction Procedure for Long Devices} 
A long $p$MOS ($W/L=\SI{10}{\micro\meter}/\SI{1}{\micro\meter}$), a long $n$MOS ($\SI{3}{\micro\meter}/\SI{1}{\micro\meter}$) and a short $n$MOS ($\SI{3}{\micro\meter}/\SI{30}{\nano\meter}$) from a commercial 28-nm bulk CMOS technology  were measured down to \SI{4.2}{\kelvin} using a Lakeshore cryogenic probe station. The transfer characteristics were recorded using a Keysight B1500A semiconductor device analyzer at $V_{DS}=\SI{10}{\milli\volt}$. 

Fig. \ref{fig:ss_p}(a) shows the measured $I_D-V_G$ curves for the long $p$MOS. The horizontal line indicates the current at which the $SS$ and $V_T$ are extracted (constant current method), respectively $I_{DS}^{(SS, \mathrm{extract})}=\SI{5e-9}{\ampere}$ and $I_{DS}^{(V_{T,\mathrm{cc}})}=\SI{3e-6}{\ampere}$. Using interpolation, the $V_G(T)$ and $V_T(T)$ data points are obtained from the experimental data at these two $I_{DS}$ values. Their difference $V_{GT}$ is then plotted in Fig. \ref{fig:ss_p}(b) [markers]. We observe that $V_{GT}$ has an approximately linear trend vs. temperature. A first-order model can be obtained for the $V_{GT}$-dependence in the model by writing two expressions for $I_{DS}$ in the high-temperature limit ($T>T_c$) [first term in (\ref{eq:IDS})]: 
\begin{eqnarray}
	I_{DS}^{(V_T^{cc})}&=&K\cdot U_T^2\cdot \left(1-e^{\frac{-V_{DS}}{U_T}}\right)\\
	I_{DS}^{(SS, \mathrm{extract})}&=& K\cdot U_T^2\cdot e^{\frac{V_{GT}}{mU_T}}\left(1-e^{\frac{-V_{DS}}{U_T}}\right)
\end{eqnarray}
Dividing both expressions, we obtain 
\begin{equation}
	V_{GT}=V_{GT0}-m\cdot U_T\cdot \ln\left(\frac{I_{DS}^{V_{T,\mathrm{cc}}}}{I_{DS}^{SS,\mathrm{extract}}}\right)
\end{equation}
where the added $V_{GT0}$ is the linearly extrapolated $V_{GT}$ at $\approx \SI{0}{\kelvin}$. The constant current method is useful here to extract $V_T$ because the $K$ (including material parameters $\mu$ and $C_{ox}$) cancels out. Fig. \ref{fig:ss_p}(c) plots the experimental $SS$ versus $I_{DS}$. The temperature trend of $SS$ is extracted at $I_{DS}=\SI{5e-9}{\ampere}$ and shown in Fig. \ref{fig:ss_p}(d). The model is then fitted to the data according to the following procedure. First, the linear $V_{GT}$ model is fit to the experimental data in Fig. \ref{fig:ss_p}(b), using $m$, $V_{GT0}$, and the two known current values. The slope factor $m$ is then further adjusted by looking at the high-temperature slope of $SS(T)$ in Fig. \ref{fig:ss_p}(d). Finally, $T_c$ (or $T_v$ for $p$MOS) is determined from the low-temperature plateau of $SS(T)$.
\begin{figure}[t]
	\centering
	\includegraphics[width=0.48\textwidth]{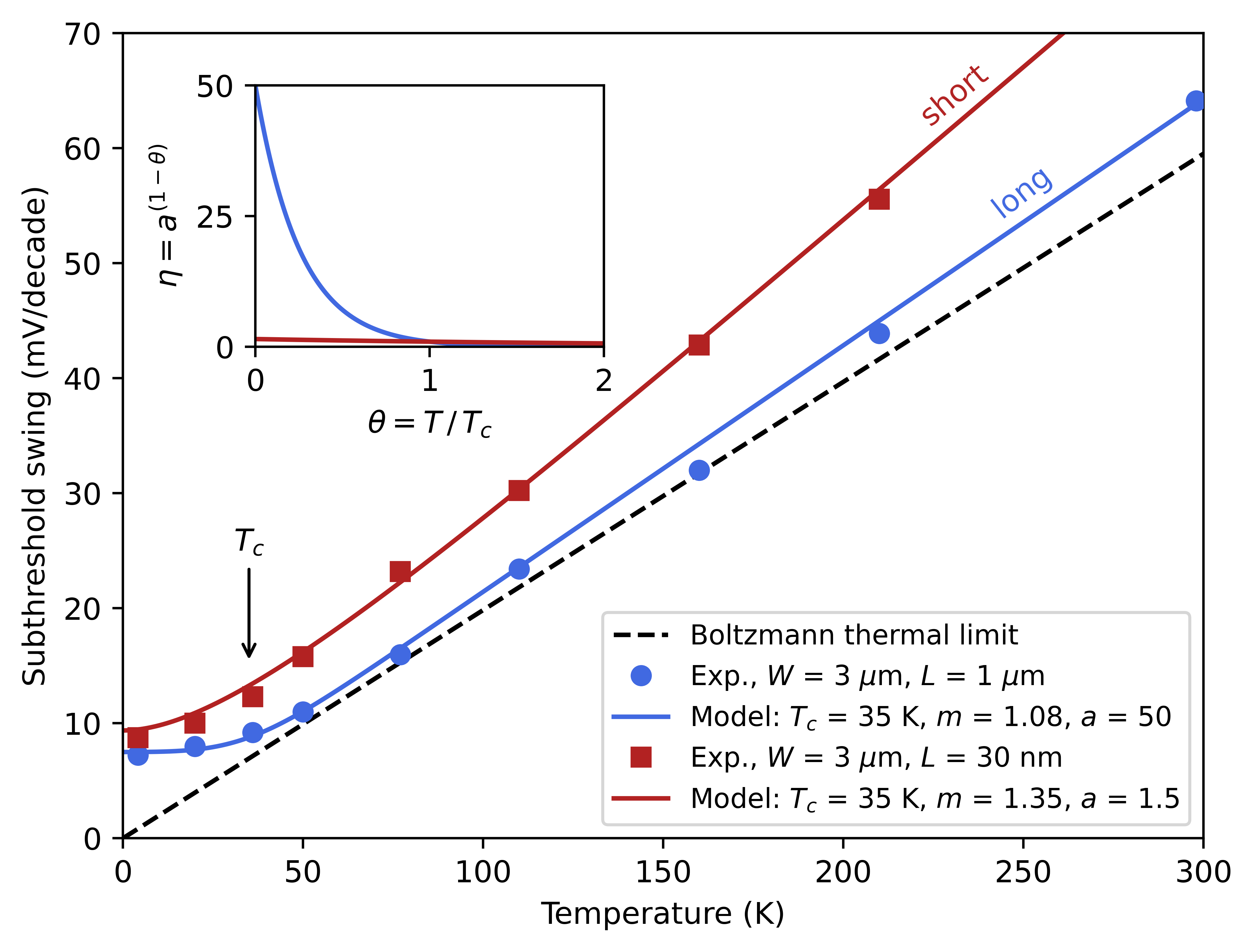}
	\vspace{-0.2cm}
	\caption{\textcolor{black}{Model (\ref{eq:ssfinal}) is calibrated to the $n$MOS devices with long and short channels. If the gate length shortens, the plateau at very low temperatures lies higher because the slope factor is larger for short devices (worse gate control). We also see a smoother roll-off around the transition region for shorter devices. Inset shows the simplified $\eta$-function from (\ref{eq:y}), where $a$ is a smoothing parameter; $a$ comes partly from the $V_{GT}$ dependence derived in the long-channel model [see Fig.\ref{fig:sensitivity}(c)] but here it is given some more freedom.}}
	\label{fig:short}
\end{figure}
\subsection{Simplified Formula for Long and Short Devices}
\textcolor{black}{As explained in Sec.\ref{sec:ss}, the model is built on long-channel assumptions. Here we give the model some additional freedom in the transition region around $T_c$ to be able to fit both long and short devices. Replacing $e^{\frac{-V_{GT}}{m\cdot U_T}}$ in (\ref{eq:y}) with a parameter $a>1$, (and $r\rightarrow 1$, i.e., neglect $V_{DS}$ in the model), we obtain}
\begin{equation}
	SS=m\cdot\left[\frac{1-a^{1-\theta}/\theta^3}{1-a^{1-\theta}/\theta^2}\right]\cdot \frac{k_BT}{q}\cdot\ln(10), 
	\label{eq:ssfinal}
\end{equation}
\textcolor{black}{which is shown in Fig. \ref{fig:short}. For long (short) devices the transition below $T_c$ to the low-temperature plateau is sharper (smoother), therefore $a$ is set to a larger (smaller) value. This causes $a^{1-\theta}/\theta^3$ and $a^{1-\theta}/\theta^2$ to dominate faster (slower) over 1 in the numerator and denominator of (\ref{eq:ssfinal}), hence $SS$ levels off faster (slower) to $m\cdot [1/\theta]\cdot (k_BT/q)\cdot \ln(10)=m\cdot (k_BT_c/q)\cdot\ln(10)$.}  

\section{Conclusion}
A closed-form and physics-based expression is derived for the saturation of $SS$ in MOSFETs based on band-tail physics. Exponential band tails can arise from Gaussian-distributed potential well depths in the channel if their binding energies are taken into account. The obtained $SS$ model showed that the Boltzmann thermal limit becomes multiplied with a new factor that depends on temperature and voltage. In the low-temperature limit, this new factor revealed the long-sought $1/T$-dependence, which has often been postulated in previous literature. We started from a more  general set of Boltzmann relations which was essential to arrive at a well-behaved transition around $T_c$. The gate-voltage dependence could partly explain the smoother roll-off around the transition temperature. The drain voltage dependence showed little impact. In the future, band-tail extraction can become a powerful tool to gauge interface quality in MOS structures for cryogenic computing. 

\appendix
Similarly, for the hole density, we have
\begin{eqnarray}
	\label{eq:p}	p &=& N_v \cdot A_v \cdot\exp\left(\frac{-\psi+\Phi_{F,p}-0.5\cdot E_g}{U_T}\right)\\&+&N_v\cdot B_v\cdot\exp\left(\frac{-\psi+\Phi_{F,p}-0.5\cdot E_g}{U_{T,v}}\right),\nonumber
\end{eqnarray}
where $A$ and $B$ remain the same as in (\ref{eq:A}), except that $\theta=T/T_v$, with $k_BT_v$ the extension of the valence-band tail.

Since the valence-band tail will have little to no influence on the $SS$ for an $n$MOSFET, we make a symmetric band-tail approximation ($T_c\approx T_v$; $N_c\approx N_v$) and assume that $A_v=B_v=1$. Furthermore, the effects of dopant freeze-out and dopant ionization can be ignored in computing the $SS$. Therefore the charge-neutrality in the $p$-type bulk of the MOSFET ($\psi=0$) is simply $p\approx N_A$, which yields two useful formulas at low and high temperatures, respectively:
\begin{equation}
	\exp\left(\frac{-0.5\cdot E_g}{U_{T,c}}\right)=\frac{N_A}{N_c}\cdot \exp\left(\frac{-\Phi_{F}}{U_{T,c}}\right),
	\label{eq:1}
\end{equation}
\begin{equation}
	\exp\left(\frac{-0.5\cdot E_g}{U_{T}}\right)=\frac{N_A}{N_c}\cdot \exp\left(\frac{-\Phi_{F}}{U_{T}}\right).
	\label{eq:2}
\end{equation}
where $\Phi_{F,p}\triangleq\Phi_F=\Phi_{F,n}-V$, \textcolor{black}{and $V$ is the channel voltage or difference in quasi-Fermi potentials}. Inserting (\ref{eq:1}) and (\ref{eq:2}) into (\ref{eq:n}), we obtain the following expression 
\begin{eqnarray}
	\label{eq:npsioverNA}\frac{n(\psi)}{N_A}&=& A_c \cdot\exp\left(\frac{\psi-2\Phi_{F}-V}{U_T}\right)\\&+&B_c\cdot\exp\left(\frac{\psi-2\Phi_{F}-V}{U_{T,c}}\right).\nonumber
\end{eqnarray}

Combining (\ref{eq:npsioverNA}) with (\ref{eq:Qi}), integrating, and applying the subthreshold relation, $V_{GS}=m\cdot \psi_s$, \textcolor{black}{and the threshold-voltage definition, $V_T\triangleq m\cdot 2\Phi_F$}, we obtain the sheet-charge density,
\begin{eqnarray}
	\label{eq:Qi2}	-Q_i&=& (m-1)\cdot C_{ox}\cdot U_T\cdot A_c\cdot e^{\frac{V_{GT}-mV}{m\cdot U_T}}\\
	&+&(m-1)\cdot C_{ox}\cdot U_{T,c}\cdot B_c\cdot e^{\frac{V_{GT}-mV}{m\cdot U_{T,c}}},\nonumber
\end{eqnarray}
where $V_{GT}=V_{GS}-V_T<0$ in subthreshold. The exponential with the largest denominator ($m\cdot U_T$ or $m\cdot U_{T,c}$) will dominate. At high $T$ ($U_T\gg U_{T,c}$ or $\theta\gg1$), this is the first exponential, and $A_c\rightarrow 1$, which recovers the standard $Q_i$ at room temperature \cite{taur}. At low $T$ ($U_T\ll U_{T,c}$ or $\theta\ll 1$), the second exponential dominates, and $B_c\rightarrow 1 \left[\,\sin(x)\approx x \right.$ for small $\left. x \, \right]$. The drift-diffusion current in the MOSFET is \cite{taur}
\begin{equation}
	I_{DS}=\mu\cdot \frac{W}{L}\int_0^{V_{DS}}-Q_i(V)\cdot dV,
	\label{eq:driftdiffusion}
\end{equation}
where $\mu$ is the channel mobility, $W$ and $L$ the width and length of the transistor, and $V_{DS}$ the drain-to-source voltage. Integrating (\ref{eq:driftdiffusion}) with $Q_i$ from (\ref{eq:Qi2}), we find a voltage- and $T$-dependent expression for the subthreshold current:
\begin{eqnarray}
	I_{DS}&=& K\cdot U_T^2\cdot A_c\cdot e^{\frac{V_{GT}}{m\cdot U_T}}\left(1-e^{\frac{-V_{DS}}{U_T}}\right)	\label{eq:IDS}	\\
	&+&K\cdot U_{T,c}^2\cdot B_c\cdot e^{\frac{V_{GT}}{m\cdot U_{T,c}}}\left(1-e^{\frac{-V_{DS}}{U_{T,c}}}\right)	,\nonumber
\end{eqnarray}
where $K=\mu\cdot \frac{W}{L}\cdot (m-1)\cdot C_{ox}$.

\bibliographystyle{ieeetran}
\bibliography{references}

\end{document}